\documentclass[12pt]{article}
\usepackage{epsfig}

\newcommand{\be}{\begin{equation}}
\newcommand{\ee}{\end{equation}}

\def\bsg{\ifmmode B\to X_s\gamma\else $B\to X_s\gamma$\fi}
\def\bsll{\ifmmode B\to X_s\ell^+\ell^-\else $B\to X_s\ell^+\ell^-$\fi}
\def\shat{\ifmmode \hat{s}\else $\hat{s}$\fi}

\newcommand{\newc}{\newcommand}

\newc{\gsim}{\lower.7ex\hbox{$\;\stackrel{\textstyle>}{\sim}\;$}}
\newc{\lsim}{\lower.7ex\hbox{$\;\stackrel{\textstyle<}{\sim}\;$}}
\newc{\ie}{{\it i.e.}}
\newc{\etal}{{\it et al.}}
\newc{\mev}{\hbox{\rm\,MeV}}
\newc{\gev}{\hbox{\rm\,GeV}}
\newc{\tev}{\hbox{\rm\,TeV}}
\newc{\xpb}{\hbox{\rm\, pb}}
\newc{\xfb}{\hbox{\rm\, fb}}

\newc{\mtop}{m_t}
\newc{\mbot}{m_b}
\newc{\mz}{M_Z}
\newc{\mw}{M_W}
\newc{\alphasmz}{\alpha_s(M_Z)}
\newc{\swsq}{\sin^2\theta_W}
\newc{\cwsq}{\cos^2\theta_W}
\newc{\tw}{\tan\theta_W}
\newc{\cw}{\cos\theta_W}
\newc{\sw}{\sin\theta_W}
\newc{\BR}{\hbox{\rm BR}}
\newc{\zbb}{Z\to b\bar}
\newc{\Gb}{\Gamma (Z\to b\bar b)}
\newc{\Gh}{\Gamma (Z\to \hbox{\rm hadrons})}
\newc{\sgn}{\mbox{sgn}}

\newlength{\myem}
\newcounter{mysubequation}[equation]

\def\beq{\begin{equation}}
\def\eeq{\end{equation}}
\def\bea{\begin{eqnarray}}
\def\eea{\end{eqnarray}}
\def\slashchar#1{\setbox0=\hbox{$#1$}           
   \dimen0=\wd0                                 
   \setbox1=\hbox{/} \dimen1=\wd1               
   \ifdim\dimen0>\dimen1                        
      \rlap{\hbox to \dimen0{\hfil/\hfil}}      
      #1                                        
   \else                                        
      \rlap{\hbox to \dimen1{\hfil$#1$\hfil}}   
      /                                         
   \fi}                                         %
\catcode`@=11
\long\def\@caption#1[#2]#3{\par\addcontentsline{\csname
  ext@#1\endcsname}{#1}{\protect\numberline{\csname
  the#1\endcsname}{\ignorespaces #2}}\begingroup
    \small
    \@parboxrestore
    \@makecaption{\csname fnum@#1\endcsname}{\ignorespaces #3}\par
  \endgroup}
\catcode`@=12

\begin{document}

\baselineskip=18pt

\setcounter{footnote}{0}
\setcounter{figure}{0}
\setcounter{table}{0}

\begin{titlepage}

\begin{center}
\vspace{1cm}

{\Large \bf  Euclidean Wormholes in String Theory}

\vspace{0.8cm}

{\bf Nima Arkani-Hamed$^1$, Jacopo Orgera$^2$, Joseph Polchinski$^3$}

\vspace{.5cm}

{\it $^1$ Jefferson Laboratory of Physics, Harvard University,\\
Cambridge, Massachusetts 02138, USA}

{\it $^2$ Department of Physics, UCSB, \\ Santa Barbara, California
93106, USA}

{\it $^3$ Kavli Institute for Theoretical Physics \\Santa Barbara,
California 93106, USA}

\end{center}
\vspace{1cm}

\begin{abstract}
We show that toroidal compactification of type II string theory to six dimensions admits axionic euclidean wormhole solutions. 
These wormholes can be inserted into $AdS_3 \times S^3 \times T^4$ backgrounds, which have a well-defined CFT dual. AdS/CFT duality then suggests that the 
wormhole solutions cannot be interpreted using $\alpha$ parameters as originally suggested by Coleman. 

\end{abstract}

\bigskip
\bigskip


\end{titlepage}

\section{Introduction}

Euclidean wormholes~\cite{GS1,LRT,Hawking} are extrema of the action in euclidean quantum gravity, 
connecting two different asymptotic regions, or arbitrarily
separated points in the same geometry. These are clearly
interesting objects. The former configurations might provide for
some imprint of other vacua in the landscape on physics in our
vacuum.  The latter configurations induce in the effective action (on scales larger than the wormhole size) bilocal operators of the
form
\begin{equation}
S_{WH} = -\frac{1}{2} \sum_{IJ} \int d^D x\, d^D y\, {\cal O}_I(x) C_{IJ} {\cal
O}_J(y)\ , \label{bilocal}
\end{equation}
where the ${\cal O}_I(x)$ are local operators with
the same quantum numbers as the mouth of the wormhole at $x$.
Naively these operators completely destroy macroscopic locality. However,
Coleman argued for a different interpretation \cite{Coleman:1988cy,GS2} noting that
\begin{equation}
e^{-S_{WH}} = \int d \alpha_I\, e^{-\frac{1}{2} \alpha_I (C^{-1})_{IJ}
\alpha_J} e^{-\int d^D x \sum_I \alpha_I {\cal O}_I(x)}\ . \label{alpha}
\end{equation}
Thus, all correlation functions are those of a local theory with the
addition of the operators $\alpha_I {\cal O}_I$ to the Lagrangian density, and with a Gaussian
weighting for the $\alpha_I$.  Note that the $\alpha_I$ are constant in space and time.  Branches of the wavefunctions with different values of the $\alpha_I$ will decohere, and a given universe can be thought of as being in a superselection sector labeled by a specific set of these parameters.  

If wormholes exist in quantum gravity and are to be interpreted a la Coleman, then they represent an intrinsic randomness of the observed constants of nature, even if the fundamental Lagrangian is completely fixed and we are in a specified vacuum.\footnote{Coleman later proposed an additional mechanism that would fix the constants~\cite{ccc}, and which would in particular set the cosmological constant to zero.  This requires additional assumptions about the interpretation of euclidean gravity, and has been criticized on various grounds.}  Thus it is important to understand whether this effect is present.  AdS/CFT duality should provide a laboratory for addressing this question within string theory, as it represents a complete description of quantum gravity within AdS boundary conditions.  There have been some studies of this question~\cite{Rey:1998yx,Maldacena:2004rf} but as yet no sharp result, because the wormholes that have been considered have fields that grow at long distance on at least one side.  Thus they cannot be inserted into AdS space without changing the boundary conditions, and their effects cannot be represented locally on each end as in eq.~(\ref{bilocal}).

It is clearly of interest to try and embed wormholes into string theory and in particular into AdS/CFT. The simplest controllable wormhole solutions, with parametrically low curvature relative to the Planck scale, are the
axionic wormholes of Giddings and Strominger \cite{GS1}.
These have yet to be embedded in string theory, for the usual reason that axions are
always accompanied by dilatonic moduli which must be taken into
account.  The simplest attempt at doing so leads to singular solutions \cite{stringwh}.
Subsequent work by Tamvakis~\cite{stringwh2} and much more recent work of Bergshoeff
et.\ al~\cite{stringwh3}  revealed wormholes solutions in string compactifications down
to $D=4$ flat spacetime. These solutions can be
embedded into euclidean $AdS_2 \times S^2$ by arranging for the wormhole size to be much less than the AdS curvature.   However, this does not yet allow us to pose a sharp paradox, because the $CFT_1$ duals remain mysterious and so it
is hard to draw conclusions.   Another strategy would be to use the recent compactifications of type IIA supergravity down to $AdS_4$ with all moduli except some axions stabilized~\cite{sham}.  However, we again do not understand the CFT duals of these models; also, it is not clear whether the moduli are sufficiently massive. 

The above results are consistent with the possibility that the moduli singularities of string wormholes always conspire to prevent a sharp confrontation with AdS/CFT duality~\cite{Rey:1998yx}.\footnote{There is a fairly large literature on string wormhole solutions whose metric is nonsingular in some frame, but where the dilaton diverges at one end so these cannot be glued into a single spacetime.  Ref.~\cite{Einhorn:2002am} discusses wormholes that are asymptotically nonsingular but have a singularity in the middle.}  However, we will find that this is not the case.

In Sec.~2 we review and generalize the constructions of wormholes in flat spacetime of Refs.~\cite{GS1, stringwh, stringwh2, stringwh3}.  For euclidean gravity coupled to scalar moduli with a general metric on moduli space, we show that non-singular wormhole 
solutions exist when there are sufficiently long timelike geodesics in moduli space, measured in Planck units.  We then check whether the simplest moduli spaces that arise from toroidal compacatification down to $D$ flat dimensions satisfy this constraint. For a simple class of geodesic trajectories, we find that while there are no solutions for $6 < D \le 10$, in $D \leq 6$ non-singular wormhole solutions exist.  The $D=6$ solutions can be embedded (again on scales smaller than the AdS curvature) into $AdS_3 \times S^3 \times T^4$, where there is a well-defined CFT dual.  

In Sec.~3 we discuss some technical subtleties related to the matter path integral.  This is not central to our main point, but is necessary in particular to resolve a puzzle regarding the single-valuedness of the fields. We also evaluate the wormhole action, taking account of surface terms that are ignored in the earlier discussion.

In Sec.~4 we discuss the implications of our result.  We draw a sharp paradox between AdS/CFT duality and the fluctuation of the coupling constants, and so argue that Coleman's effect must not be present in string theory.   We also discuss alternate interpretations.

\section{Wormhole Solutions}
\subsection{Generalities}

The simplest setting for string wormhole solutions~\cite{GS1, stringwh} is a euclidean theory in $D$ dimensions, with gravity, a scalar, and a $(D-1)$-form field strength.  The action is
\begin{equation}
S_F = \frac{1}{2\kappa^2} \int d^D x \sqrt{g} \left( -{\cal R} + \frac{1}{2} \partial_\mu
\varphi \partial^\mu \varphi + \frac{1}{2(D-1)!} e^{\beta \varphi} F_{\mu \cdots \nu} F^{\mu\cdots\nu}
\right) \ , \label{gff}
\end{equation}
up to a surface term that we leave for the next section.
The equations of motion for this action are the same as for
\begin{equation}
S_A = \frac{1}{2\kappa^2} \int d^D x \sqrt{g} \left( -{\cal R} + \frac{1}{2} \partial_\mu
\varphi \partial^\mu \varphi - \frac{1}{2} e^{-\beta \varphi} \partial_\mu
A \partial^\mu A 
\right) \ , \label{gfa}
\end{equation}
with the form field replaced by an `axion' via
\begin{equation}
F_{\mu \cdots \nu} = \epsilon_{\mu \cdots \nu \lambda} e^{-\beta \varphi} \partial^\lambda A\ .
\label{hodge}
\end{equation}
We will refer to Eqs.~(\ref{gff}, \ref{gfa}) as the flux form and the axion form respectively.
Note that the Lagrangian density is not invariant, and that for real $F_{\mu \cdots \nu}$ the $A$ kinetic term is negative.  For now we use the scalar version simply as a means of encoding the equations of motion.  Subtleties such as the single-valuedness of $A$, and all surface terms in the action, will be left for the next section.

We now generalize to an arbitrary euclidean theory of gravity coupled to massless scalars
$\phi_I$ in $D$ dimensions.  The two-derivative action is
\begin{equation}
S_A =  \frac{1}{2\kappa^2} \int d^D x \sqrt{g} \left( -{\cal R} + \frac{1}{2} G_{IJ}(\phi) \partial_\mu
\phi_I \partial^\mu \phi_J \right) \ . \label{modact}
\end{equation}
As we see in the example above, for the euclidean Lagrangians coming
from compactifications of string theory, the metric $G_{IJ}$ does
not have a definite signature --- in particular the axionic scalars have
the ``wrong" sign kinetic terms.  Indeed these wrong signs are
crucial for the wormhole solutions to exist.

We are interested in spherically symmetric solutions of the form
\begin{equation}
ds^2 = N^2(r) dr^2 + a^2(r) d \Omega^2_{D-1}\ , \quad \phi_I =
\phi_I(r)\ .
\end{equation}
Plugging this Ansatz into the action we have
\begin{equation}
S_A  = \frac{V_{D-1}}{2\kappa^2 } \int dr\, N a^{D-1} \left[(D-1)(D-2)
\left(-\frac{a^{\prime 2}}{N^2 a^2} - \frac{1}{a^2} \right) + \frac{1}{2N^2} G_{IJ}
\phi^{I\prime} \phi^{J\prime} \right]\ , \label{Anact}
\end{equation}
where primes will always denote derivatives with respect to $r$.  Here we have neglected surface terms that don't affect the equations of motion,
we will take them into account in evaluating the action in the next section. 

Varying $N$, and then choosing the gauge $N(r) = 1$, gives the constraint
\begin{equation}
\frac{a^{\prime 2}}{a^2} - \frac{1}{a^2} - \frac{G_{IJ}
\phi^{I\prime} \phi^{J\prime}}{2(D-1)(D-2)} = 0 \ .
\end{equation}

The equation of motion for the scalars is
\begin{equation}
(a^{D-1} G_{IJ} \phi^{J\prime})^\prime - \frac{1}{2} a^{D-1} G_{JK,I} \phi^{J\prime}
\phi^{K\prime} = 0\ .
\end{equation}
If we define ${dr}/{a^{D-1}} = d \tau$, this becomes the
geodesic equation in the metric $G_{IJ}$,
\begin{equation}
\frac{d^2 \phi^I}{d \tau^2} + \Gamma^I \!_{JK} \frac{d \phi^J}{d \tau}
\frac{d \phi^K}{d \tau} = 0 \ .
\end{equation}
In particular these means that the invariant length $G_{IJ} (d
\phi^I/d \tau)(d \phi^J/d \tau)$ is a constant of the motion, or
equivalently that
\begin{equation}
G_{IJ} \phi^{I\prime} \phi^{J\prime} = \frac{C}{a^{2D - 2}}\ .
\end{equation}
Inserting this into the equation for $a^\prime$, we have
\begin{equation}
a^{\prime 2} -1 - \frac{C}{2(D-1)(D-2) a^{2D - 4}} = 0\ ,
\end{equation}
which is the same as energy conservation for a newtonian particle with
effective potential 
\begin{equation}
V_{\rm eff}(a) = -1 - \frac{C}{2(D - 1)(D-2) a^{2D - 4}} \ .
\end{equation}
The nature of the solution then depends on whether the geodesic
motion on moduli space is spacelike ($C > 0$), null ($C = 0$) or
timelike $(C < 0)$. For $C > 0$, the potential goes to $-\infty$ as
$a \to 0$, so $a^\prime$ must diverge and the solution is singular.
For $C=0$, we have $a(r) = r$, so the metric is that of flat space
and the scalar solution corresponds to an extremal D-instanton.
But clearly for 
\begin{equation}
C \equiv - 2(D - 1)(D-2) a_0^{2D - 4} < 0\ ,
\end{equation}
wormhole solutions are
possible with $a(r) \to r^2$ as $r \to \pm \infty$, and with minimum
value $a(r=0) = a_0$.

We see that the scalars are just travelling along a timelike geodesic
in moduli space, but there is a constraint in order to be able
to find a solution. Denote the value of the moduli
at the spacetime boundaries as $\phi_{\pm \infty}$ and the value at the neck
by $\phi_0$. Now, the timelike distance between $\phi_{-\infty}$ and
$\phi_\infty$ along the geodesic is
\begin{eqnarray}
d[\phi_{-\infty}, \phi_\infty] = 2d[\phi_{-\infty}, \phi_0] &=& 2\int_{-\infty}^0 dr
\frac{|C|^{1/2} }{a(r)^{D-1}} \nonumber \\
&=& 2\sqrt{2 (D-1)(D-2)} \int_1^\infty \frac{d \hat{a}}{\hat{a} \sqrt{ \hat a^{2D-4}  - 1}}
\nonumber\\
 &=&
{\pi}\sqrt{\frac{2(D-1)}{D-2}}\ , \label{geo}
\end{eqnarray}
where we substituted $dr \to da/a^{\prime}(r)$, used the equation
of motion for $a^\prime$, and introduced the dimensionless variable  $\hat{a} = a/a_0$. Thus, in
order to be able to find a wormhole solution with the moduli bounded, we must be able to
find a compact timelike geodesic  at least as long as~(\ref{geo}) in the scalar moduli
space.  
In practice we will identify noncompact timelike geodesics, whose length must be strictly greater than (\ref{geo}),
\begin{equation}
d^2 > 2 \pi^2 \frac{D-1}{D-2} \ , \label{dmax}
\end{equation}
so that we can restrict to a compact piece satisfying~(\ref{geo}).
This is measured in Planck units, $2\kappa^2 = 1$.

It is trivial to generalize this analysis to a case with negative
cosmological constant, so that the asymptotic spaces are AdS spaces
with curvature scale $L$.  This is just an aside; it is not directly relevant to the case we will be interested in, instantons localized on AdS$\times$S spaces.
The cosmological constant simply adds a piece $- a^2/L^2$ to
the effective potential $V(a)$, and the solutions asymptote to $a(r)
\to e^{|r|/L}$ as $r \to \infty$.  Following the same steps as before, the length of the geodesic is
becomes
\begin{equation}
d[\phi_{-\infty},\phi_0] = \sqrt{2 (D-1)(D-2)} \int_1^\infty \frac{d
\hat{a}}{\hat{a} \sqrt{ \hat a^{2D-4} ( \hat a^2 + \hat L^2)/(1 + \hat L^2)  - 1}}\ ,
\label{ads}
\end{equation}
where $\hat a = a/a_0$, $\hat L = L/a_0$, and $a_0$ is again the turning radius.
This is strictly less than the flat space integral, so the necessary geodesic is shorter (a weaker condition).  Clearly for small
wormholes relative to the AdS scale $a_0/L \ll 1$, the bound on the
length is the same as before up to corrections of order
$O(a_0^2/L^2)$.

The condition of having a long enough timelike geodesic in moduli
space is trivially satisfied for the axion-gravity system without a dilaton~\cite{GS1}, where the
moduli space is one-dimensional and the metric 
$
ds^2 = - d A^2
$
is
timelike thanks to the ``wrong" sign of
the axion kinetic term. However the moduli spaces we get from simple
compactifications of string theory have both space-like and
time-like directions, associated with axions and dilatons. In particular, 
these Lorentzian moduli spaces have horizons that limit the length of timelike geodesics.  For instance, consider the axion-dilaton
system in type IIB string theory in $D=10$. The moduli space metric
(with the wrong sign for the axion) is
\begin{equation}
d s^2 = d \varphi^2 - e^{2 \varphi} d A^2 \ ,
\end{equation}
corresponding to $\beta = -2$ in the action~(\ref{gff}, \ref{gfa}). 
This is just minus the metric of a causal patch of $(1+1)$-dimensional de Sitter space with
unit dS radius, so we are interested in spatial geodesics in this patch.  As is familiar, because of the
presence of the dS horizon there is a maximum separation between
spacelike separated points beyond which no connect can
connect them, so there is a maximum length for spatial geodesics.

This can be found by an easy direct computation: there is a noncompact timelike geodesic $e^{\varphi} = \cos \tau$, $A = \tan\tau$, whose length is $\pi$.  We can also
get the answer indirectly, by remembering that the Wick
rotation of a causal patch of dS is half of a ball $S^2$, so the
maximum separation angle, and the maximum length on the unit hemisphere, is $\pi$.\footnote{There is a geodesic of length $2\pi$ running around the edge, but there is no slicing such that this continues back to a Lorentzian geodesic.}
There is no $D$ for which this exceeds ${\pi} \sqrt{\frac{D-1}{D-2}}$ (the case $D=1$ is unphysical), so there are no wormhole solutions.

For general $\beta$, the metric can be written
\begin{equation}
ds^2 =  d \varphi^2 - e^{-\beta \varphi} d A^2 \equiv \frac{4}{\beta^2} (d\tilde \varphi^2 - e^{2 \tilde \varphi} d \tilde A^2)\ .
\end{equation}
The longest geodesic is now $2\pi/|\beta|$, and so the condition~(\ref{dmax}) for a wormhole solution becomes~\cite{GS1,stringwh,Gutperle:2002km}
\begin{equation}
\frac{1}{\beta^2} > \frac{D-1}{2(D-2)}\ . \label{bcrit}
\end{equation}
For reference we give the most general long geodesic for given $\beta$, whose four parameters $\varphi_0, A_0, \tau_0, q$ can be obtained by a general $SL(2,R)$ transformation together with a rescaling of the affine parameter, which corresponds to scaling the wormhole charge (a shift of the affine parameter has the same effect as one of the $SL(2,R)$ generators):
\begin{equation}
e^{-\beta \varphi/2} = e^{- \beta \varphi_0/2} \cos q(\tau - \tau_0)\ ,\quad A = A_0 - \frac{2}{\beta}e^{\beta \varphi_0/2} \tan q(\tau - \tau_0)\ , \label{gensol}
\end{equation}
The length, for $-\pi/2 < q\tau < \pi/2$, is always $2\pi/|\beta|$.  

\subsection{Wormholes in $D = 4, 6$}

For heterotic or type II superstring theory compactified to $D=4$ on a Calabi-Yau manifold or $T^6$, the effective action for the four-dimensional dilaton and axion $(\Phi_4, B_{\mu\nu})$ plus the internal dilaton and axion $(\sigma,A)$ is
\begin{equation}
S = \frac{1}{2\kappa^2} \int d^4 x \sqrt{g} \left( -{\cal R} + 2 \partial_\mu \Phi_4 \partial^\mu \Phi_4 + 6 \partial_\mu \sigma \partial^\mu \sigma
+ \frac{1}{12} e^{-4\Phi_4} H_{\mu \nu\rho} H^{\mu\nu\rho}
 - \frac{1}{2} e^{-4\sigma} \partial_\mu
A \partial^\mu A 
\right) \ . \label{d4act}
\end{equation}
Upon rescaling the fields, this is two copies of the system~(\ref{gff}, \ref{gfa}), the four-dimensional one having $\beta^{-2} = \frac{1}{4}$ and the internal one having $\beta^{-2} = \frac{3}{4}$~\cite{stringwh}.  The condition~(\ref{bcrit}) becomes~$\beta^{-2} > \frac{3}{4}$ so both of these wormholes are singular.\footnote{For the critical case $\beta^{-2} = \frac{3}{4}$, the fields blow up only at infinity, and it is possible that an extension of the analysis~(\ref{ads}) to $AdS_2 \times S^2$ would give a nonsingular wormhole.  However, to make the sharpest paradox we would like to be able to take the wormholes small compared to the AdS radius, so we insist that they be nonsingular even in flat space.}  However, a simple observation~\cite{stringwh2, stringwh3} allows the construction of nonsingular wormholes in this theory.  That is, if we consider a solution with {\it both} axion charges, then the relevant moduli space is the product of the two separate spaces, and the longest timelike geodesic would be the `diagonal' in the two spaces.  The condition~(\ref{bcrit}) now becomes
\begin{equation}
\sum_i \frac{1}{\beta_i^2} > \frac{D-1}{2(D-2)}\ , \label{bicrit}
\end{equation}
which is comfortably satisfied in the theory~(\ref{d4act}).  Thus euclidean wormhole solutions do exist in string theory~\cite{stringwh2, stringwh3}.

Turning to the IIB theory on $T^4$, for which we have a good CFT dual, the most closely analogous action would involve the four-dimensional and internal dilatons, as well as the noncompact RR five-form field strength and internal axion.  The reduced action is
\begin{equation}
S = \frac{1}{2\kappa^2} \int d^6 x \sqrt{g} \left( -{\cal R} + \partial_\mu \Phi_6 \partial^\mu \Phi_6 + 4 \partial_\mu \sigma \partial^\mu \sigma
+ \frac{1}{12} e^{- 2\Phi_6 + 4\sigma } F_{\mu\cdots\rho} F^{\mu\cdots\rho}
 - \frac{1}{2} e^{-4\sigma} \partial_\mu
A \partial^\mu A 
\right) \ . \label{d6act}
\end{equation}
The internal dilaton-axion system has $\beta^{-2} = \frac{1}{2}$.   Exciting the five-form sources the linear combination $\Phi_4 = -2\sigma$, giving $\beta^{-2} = \frac{1}{4}$.  Neither of these exceeds the necessary value $\frac{5}{8}$.  Also, in this case we cannot simply combine the two systems diagonally as in Eq.~(\ref{bicrit}) because the moduli space is not a product: the dilatons mix, and so do the axions (through Chern-Simons couplings); we do not know if there is a sufficiently long geodesic in this space.

However, a simple trick allows us to find nonsingular wormholes in a different way.  Consider just the internal fields, and regard the $T^4$ as $T^2 \times T^2$.  We now have a product space where each piece has $\beta^{-2} = \frac{1}{4}$, so that the diagonal geodesic has length-squared $\frac{1}{4} + \frac{1}{4}$,  reproducing the result for the internal dilaton-axion of $T^4$.  Now, it is familiar that for compactification on $T^2$ we can identify {\it two} decoupled dilaton-axion systems, where the first is from the dilaton-axion on the $T^2$ and second comes from the complex structure of the $T^2$.  A $Z_2$ $T$-duality interchanges these so they must each have $\beta^{-2} = \frac{1}{4}$.  Summing over the four separate factors from the two $T^2$s, the left-hand-side of Eq.~(\ref{bicrit}) is  $1 >\frac{5}{8}$, and so there are nonsingular wormholes with six noncompact dimensions. Thus we will be able to frame a sharp paradox with AdS/CFT.

The axions for the solution just described come from $g_{67}$, $B_{67}$, $g_{89}$, and $B_{89}$.  We can also construct this solution in various dual forms.  For example, by taking the $S$-dual, and then the $T$-dual on the 7- and 9-axes, we obtain instead the axions $B_{67}$, $C_{69}$, $B_{89}$, and $C_{78}$.  The reduced action is
\begin{eqnarray}
S &=& \frac{1}{2\kappa^2} \int d^6 x \sqrt{g} \left( -{\cal R} + \partial_\mu \Phi_6 \partial^\mu \Phi_6 + \sum_{m = 6}^9 \partial_\mu \sigma_m \partial^\mu \sigma_m
\right.\nonumber\\
&&\qquad
 -\frac{1}{2} e^{-2\sigma_6 - 2\sigma_7} \partial_\mu B_{67} \partial^\mu B_{67}
-\frac{1}{2} e^{-2\sigma_8 - 2\sigma_9} \partial_\mu B_{89} \partial^\mu B_{89}\nonumber\\
&&\qquad \left.
-\frac{1}{2} e^{2\Phi_6 - \sigma_6 + \sigma_7 + \sigma_8 - \sigma_9} \partial_\mu C_{69} \partial^\mu C_{69}
-\frac{1}{2} e^{2\Phi_6 + \sigma_6 - \sigma_7 - \sigma_8 + \sigma_9} \partial_\mu C_{78} \partial^\mu C_{78}
\right) \ . \label{d6act2}
\end{eqnarray}
Note that the axions couple to orthogonal combinations of moduli, and that the normalization corresponds to $\beta^{-2} = \frac{1}{4}$ for each.
For concreteness, we will focus on this example in the following discussion.

The $D=6$ wormhole solution can also be understood in terms of the $SO(5,5)/SO(5)\times SO(5)$ geometry of the moduli space.   For the numerator group we have $SO(5,5) \supset SO(4,4) \supset SO(2,2)^2 = SO(2,1)^4$; the last step is parallel to the familiar $SO(4) = SO(3)\times SO(3)$.
The intersection of the denominator group with the $SO(2,1)^4$ is $SO(2)^4$.  Thus we obtain four copies of the dilaton-axion system.  The construction with axions $g_{67}$, $B_{67}$, $g_{89}$, and $B_{89}$ corresponds to $SO(4,4)$ and $SO(2,2)\times SO(2,2)$ being the real versions of the $T$-duality groups of $T^4$ and $T^2 \times T^2$ respectively.  The axions  $B_{67}$, $C_{69}$, $B_{89}$, and $C_{78}$ lie in a $U$-dual $SO(4,4)$.  In $D=7$ the numerator group is $SL(5,R)$ and only contains two copies of  $SO(2,1)$, which is not enough for our construction.  It is possible that there are longer geodesics not lying in a product of $SO(2,1)/SO(2)$ factors, but we have not been able to find any.

\section{Technicalities}

\subsection{Path integral subtleties}

We begin with a discussion of the structure of the euclidean action, in particular the peculiar ``wrong" sign kinetic terms in the axion form 
of the action compared to the ``normal" sign kinetic term for the ``flux" form. From a variety of perspectives, it is nice to understand what is going on 
in axionic language; for instance because winding $D$ and $F$ strings have a natural local coupling to axions. 
Much of the discussion below is a review of the arguments
of Coleman and Lee \cite{Coleman:1989ky}. We then compute the action for the wormhole solutions found in the previous section. Interestingly, 
we find that this action is always {\it smaller} than that of a pair of D-instantons with the same charges at the two mouths of the wormhole.

The relevant issues all arise in a very simple and familiar toy setting: the quantum mechanics of a non-relativistic particle moving on a 
circle of unit radius, with euclidean action $S = \int dt \left(\frac{1}{2} \dot{\theta}^2 + V(\theta)\right)$. Here $\theta$ is the analog of our axions. With vanishing potential, 
there is shift symmetry on $\theta$ with associated charge $Q$ --- the particle momentum. We include the possibility of a small $V(\theta)$ to model 
the tiny shift-symmetry violating effects we also expect in our axion example. Now consider the euclidean transition 
amplitude 
\begin{equation}
G_{\theta_f,\theta_i}(T) = \langle \theta_f | e^{-H T} | \theta_i \rangle
\end{equation}
Let us start with $V = 0$.  In this trivial case we know the spectrum exactly and 
\begin{equation}
G_{\theta_f, \theta_i}(T) = \sum_Q e^{-Q^2 T/2} e^{i Q (\theta_f - \theta_i)}
\end{equation}
Clearly, for large $T$, this expression has the form of a semiclassical expansion, with increasingly 
exponentially suppressed contributions from larger charge sectors; indeed this sum 
is the direct analog of the ``flux" presentation of the action. On the other hand, a straightforward evaluation of the path integral 
representation of $G_{\theta_f, \theta_i}(T)$ does not yield this semiclassical expansion. Indeed, the saddle points of 
the euclidean path integral with paths starting at $\theta_i$ and ending at $\theta_f$ in euclidean time $T$ are ones that wind around the 
circle $m$ times, so $\theta(\tau) = \theta_i + (\theta_f - \theta_i + 2 \pi m) \tau/T$, 
with euclidean action $(\theta_f - \theta_i + 2 \pi m)^2/2T$. Thus 
\begin{equation}
G_{\theta_f, \theta_i} (T) = \sum_m e^{-{(\theta_f - \theta_i + 2 \pi m)^2}/{2T}}  
\end{equation}
Of course this expression for $G$ is the same as the earlier one by Poisson resummation (modulo a prefactor form the determinant, which we have omitted), but for large $T$ the direct evaluation of the 
euclidean path integral gives a very bad expansion. It would be nice to extract the good semiclassical expansion directly from a $\theta$ 
path integral. To wit, let us look at
the euclidean transition amplitude, not between eigenstates of $\theta$ but between momentum eigenstates: 
\begin{equation}
G_{Q^\prime, Q}(T) = \langle Q^\prime | e^{-H T} | Q \rangle \ ,\quad G_{\theta_f, \theta_i}(T) =\sum_{Q^\prime, Q} 
e^{i(Q^\prime \theta_f - Q \theta_i )} G_{Q^\prime,Q}(T)
\end{equation}
Clearly for $V(\theta) = 0$, charge is conserved and $G_{Q^\prime, Q}$ is diagonal, while for small $V(\theta)$ there will be small off-diagonal pieces. Now, since $|Q \rangle = \int d \theta\, e^{i Q \theta} | \theta \rangle$, there is a simple path integral representation for $G_{Q^\prime,Q}(T)$:
\begin{equation}
G_{Q^\prime,Q}(T) = \int_{\rm free} {\cal D} \theta \, e^{-\tilde S(\theta)}\ ,\quad \tilde S(\theta) = {S} (\theta) - i [Q \theta(0) - Q^\prime \theta(T)]\ ,
\end{equation}
where the first term is the usual euclidean action ${S}(\theta) 
= \int_0^T \left(\frac{1}{2} \dot{\theta}^2 + V(\theta)\right)$ and the second term is a boundary action.  Note that the boundary values of $\theta $ are not fixed but free (integrated over) in this path integral.  To belabor the obvious --- if we are interested in $G_{\theta_f,\theta_i}$, the boundary values $\theta_f,\theta_i$ enter only in the fourier transform from $G_{Q^\prime,Q}$ to $G_{\theta_f, \theta_i}$, and have {\it nothing} to do with the (unfixed!) boundary values in the $\theta $ path integral.  

Now, the saddle points contributing to the $\theta $ path integral are easily determined. The equation of motion for $\theta $ is the usual euclidean 
one $\ddot{\theta} - V^\prime(\theta) = 0$, but there is also a boundary variation which leads to 
\begin{equation} 
\dot{\theta}(T) =- i Q^\prime, \, \, \, \dot{\theta}(0) = - i Q \ .
\end{equation}
Clearly in general these equations have complex solutions. For the special case $V(\theta) = 0$, 
as expected there is only a solution for 
$Q = Q^\prime$ which is $\dot{\theta} = i Q$. Alternately, we could define $\theta = - i A$; then $A$ would have the ``wrong" sign kinetic term
 but a real solution. Either way, on the solution, the action is $\tilde S = + Q^2 T/2$, and reproduces the nice semiclassical expansion for large $T$. 
 
We can consider a more interesting example, a particle moving in a central potential in two dimensions with euclidean action $S = \int dt \left(\frac{1}{2} 
\dot{r}^2 + \frac{1}{2} r^2 \dot{\theta}^2 + V(r) \right)$.  There is still a charge associated with the shift symmetry on 
$\theta$, the angular momentum, and we can still find a semiclassical expansion along the lines above. Clearly while $\theta $ in this case will be imaginary, $r$ will not be; equivalently, we flip the kinetic term for $\theta$ and not $r$ in evaluating the saddle point action with fixed angular momentum. In both cases, the kinetic term of the variable conjugate to the conserved charge is flipped; note that the action is positive, however, 
and can be obtained by inserting the solution of the equations of motion with the wrong sign kinetic term back into the 
original euclidean action with usual sign kinetic terms.

This story generalizes trivially to our wormhole example. In all the cases of interest in string theory, there is a set of co-ordinates on moduli space where the moduli naturally 
group themselves into dilatons $\varphi_a$ and axions $A_i$, with metric ${G}_{IJ} d \phi_I d \phi_J = G_{ab}(\varphi) d \varphi_a d\varphi_b + \sum_i F_i(\varphi) dA_i^2$. There are charges $Q_i$ 
associated with the shift symmetries on the $A_i$. The wormhole solutions correspond to euclidean transition amplitudes with fixed $Q_i$, and can be obtained from an action where the kinetic 
terms for the $A_i$ are flipped, yielding the metric $G_{IJ}$ we considered in the previous section. The wormhole solution follows from varying the action 
\begin{equation}
S = \frac{1}{2 \kappa^2} \int d^D x \sqrt{g} \left( -{\cal R} + \frac{1}{2} G_{IJ} \partial_\mu \phi_I \partial^\mu \phi_J \right)
\end{equation}
In order to get a good semiclassical expansion, we must insert projections onto definite $Q$ at the wormhole ends, as above, and then on equations of motion the semiclassical action is given by $\tilde S$, with the signs of the axion kinetic terms flipped.
For transitions between asymptotic regions at points $(\varphi_a, A_i)$ and $(\varphi_a^\prime, A_i^\prime)$ in the moduli space, the wormhole 
solution need only interpolate between $\varphi_a, \varphi_a^\prime$, with no regard for the the dependence on $A_i,A^\prime_i$ --- the latter are discontinuous at the projection operators.

In order for this semiclassical approximation to be valid, the wormhole should have large action. 
The wormhole carries the same charges as wrapped F- and D-string instantons, so we need 
\begin{equation}
L^2 /\alpha^\prime \gg 1\ ,\quad L^2 /\alpha^\prime g_{\rm s} \gg 1\ , \label{bigtor}
\end{equation}
where $L$ is the size of the $T^4$, assumed to be roughly isotropic.  Winding F- and D-string states couple to the axions.  In the regime~(\ref{bigtor}), these are massive compared to other scales and can be integrated out, giving rise to small breaking of the shift symmetry on the $A_i$.\footnote{R.~Myers points out that due to the complex value for the axion field in the wormhole solution, the masses-squared of these winding states may sometimes acquire a {\it negative} real part.  This should not have any effect, since it should make sense to integrate out the winding states first and then continue to complex $A_i$.  One could investigate this by adding a massive winding state field to the particle model above - it cannot have a large effect on the amplitude.}
Also, in order for the supergravity description of the wormhole to hold, the throat must be large in string units.  This implies at least one of $Q_{\rm R} g_{\rm s}$ and $Q_{\rm NS} g^2_{\rm s}$ must be large, where $Q_{\rm R,NS}$ are the charges flowing through the wormhole throat.

\subsection{The wormhole action}

In our toy example, the action $S$ corresponds to the axion action $S_A$, Eq.~(\ref{gfa}), while the additional surface term in $\tilde S$ is equivalent, upon use of the equation of motion, to flipping the sign of the axion action as in $S_F$, Eq.~(\ref{gff}).  Following the previous discussion, we use the latter in evaluating the semiclassical action:
\begin{equation}
S_F = S_A + \frac{1}{2\kappa^2} \int d^D x \sqrt{g} \sum_i F_i(\varphi) \partial_\mu A_i \partial^\mu A_i 
\ ,\ \  {S}_A =  \frac{1}{2\kappa^2}\int d^D x \sqrt{g} \left(-{\cal R} + \frac{1}{2} {\cal G}_{IJ} \partial \phi_I \partial \phi_J \right) \ .
\end{equation}
In addition, the gravitational action requires a surface term involving the extrinsic curvature of the boundary minus the extrinsic curvature of the boundary embedded in flat spacetime~\cite{Hawk}.
However, this vanishes for euclidean wormhole solutions, because these approach flat spacetime sufficiently rapidly at infinity~\cite{GS1}. 

Now, very generally $S_A$ vanishes on equations of motion, since the trace of the Einstein equation immediately implies 
${\cal R} = \frac{1}{2} G_{IJ} \partial \phi_I 
\partial \phi_J$.  Using our Ansatz for the metric and scalar solution, the second term and hence the wormhole action is 
\begin{equation} 
\frac{2\kappa^2 }{V_{D-1}} {S}_F(Q_i) = \int dr\, a^{D-1} \sum_i F_i(\varphi) A_i^{\prime 2} = \int dr\, Q_i A_i^\prime = \sum _i Q_i \Delta A_i\ . \label{eomact}
\end{equation}
Here $Q_i = a^{D-1} F_i(\varphi) A_i^\prime$ is the $i$'th conserved charge,
and $\Delta A_i$ are the changes from one end of the wormhole to the other. For the solution~(\ref{gensol}),
\begin{equation}
Q_i = -\frac{2}{\beta_i}  e^{-\beta_i \varphi_{i0}/2} q_i\ ; \quad
\Delta A_i = -\frac{4}{\beta_i} e^{\beta_i \varphi_{i0}/2} \tan q_i \tau_\infty
= -\frac{4}{\beta_i} e^{\beta_i \varphi_{i\infty}/2} \sin q_i \tau_\infty\  ,\label{deltqa}
\end{equation}
For our $D=6$ solution, each $\beta_i = 2$ and the condition~(\ref{geo}) becomes
\begin{equation}
\frac{1}{4} \sum_{i=1}^4 \left(\frac{2}{\pi}q_i \tau_\infty\right)^2 = \frac{5}{8}\ .
\end{equation}
The parameter $q_i\tau_\infty$ must be less than ${\pi}/{2}$ for all $i$ in order to have a nonsingular solution; this allows some region of parameter space, corresponding to different ratios the four charges. 

Ref.~\cite{GS1} also considers a topological term $\gamma$ in the action, proporional to the Euler number of the wormhole.  This is the analog of the string coupling constant in the world-sheet expansion.  For Calabi-Yau compactification such a term would descend from a ten-dimensional Euler number term~\cite{GS1}, which conceivably could be present with an arbitrary coefficient.  However, our compactification has a toroidal factor, so the ten-dimensional Euler number is zero.  A four-dimensional Euler number term might also be produced by string and quantum corrections, but we are assuming that the wormhole throat is large so that such higher derivative corrections are small.

For comparison, let us note that for supersymmetric instantons the geodesic is null~\cite{Gibbons:1995vg,stringwh3}, so that $dA_i = \pm(2/\beta_i) d(e^{\beta_i\varphi_i/2})$.  Thus, $\Delta A_i = \pm(2/\beta_i) \Delta(e^{\beta_i\varphi_i/2})$, where we are now referring to the change between the asymptotic region and the instanton center.
In fact, $e^{\beta_i\varphi_i/2}$ vanishes at the core, so we can write for a supersymmetric instanton
\begin{equation}
\frac{2\kappa^2 {\cal S}(Q)}{\Omega_{D-1}} = \sum_i \left|\frac{ 2  Q_i}{\beta_i} \right|e^{\beta_i\varphi_{i\infty}/2}\ .
\end{equation}
Since $|\sin  q_i \tau_\infty|$ is less than one, the wormhole action~(\ref{eomact},$\,$\ref{deltqa}) is strictly less than that of a collection of supersymmetric instantons of the same total charge in the place of the two ends of the throat.  This is a curious result: for a particle state it would correspond to violation of the BPS bound, but for an instanton there appears to be no sharp conflict with supersymmetry. 

There has been an interesting related observation in Ref.~\cite{BCPVV}, that the wormhole would map to an impossible gauge theory configuration, in which $(F - \tilde F)^2$ would have to be negative.  That is, the BPS-violating bulk instanton maps to a BPS-violating boundary instanton, and there the action has positivity properties that forbid this.

\section{Discussion}

Now let us formulate a sharp paradox.  The basic idea is that the ends of a wormhole can be arbitrarily separated in time, so that amplitudes will not satisfy cluster decomposition, whereas the dual gauge theory has local time evolution and so will satisfy cluster decomposition.  Ref.~\cite{Rey:1998yx} gave similar arguments to the effect that AdS/CFT duality is inconsistent with $\alpha$ parameters.

The CFT dual to IIB string theory on $AdS_3 \times S^3 \times T^4$ is given by the infrared limit of the $D=2$, ${\cal N}=4$, supersymmetric gauge theory with gauge group $SU(Q_1) \times SU(Q_5)$~\cite{juan}.  If one avoids special points on the moduli space of the $T^4$~\cite{Seiberg:1999xz}, the scalar potential for the gauge theory on a circle increases in all directions, so the spectrum should be discrete.  

The size of the $T^4$ is of  order $(Q_1/Q_5)^{1/4}$ in string units, so we need $Q_1 \gg Q_5, Q_5 g_{\rm s}^2$ in order that the conditions~(\ref{bigtor}) for the semiclassical expansion be valid.  The radii of the $AdS_3 \times S^3$ are of order $(g_{\rm s} Q_5)^{1/2}$ in string units, so we need $g_{\rm s} Q_5 \gg 1$ in order that these radii be large compared to the string scale.  If these conditions are satisfied then we can arrange the wormhole charges so that the throat is large compared to the string scale and small compared to the AdS radius.  We can take $Q_1$ and $Q_5$ to be large but finite, and this is a superrenormalizable theory, so there should be no subtlety in regarding this as an ordinary quantum mechanical system.

For nonsingular wormholes the fields fall off as $1/r^4$ in flat spacetime (like the Coulomb Green's function).  At longer distance this will go over to the Coulomb Green's function for $AdS_3 \times S^3$,
\begin{equation}
G(\tau,\theta) = \frac{1}{4 ( \cosh \tau -  \cos\theta)^2}\ , \label{adscoul}
\end{equation}
where $\tau$ and $\theta$ are the distance along $AdS_3$ and $S^3$ respectively.  This is normalizable at infinity, so these wormholes, if present, would represent effects described by the original CFT rather than a perturbation of the CFT~\cite{GKPholo,Wholo}.\footnote{Depending on which $U$-dual form of the wormhole solution we use, there may be Chern-Simons couplings of the axions.  These give rise to AdS masses, and so the Green's function falls off faster.}
The wormhole ends interact through their long-range fields.  The wormhole solution is thus not exact --- its action depends on the separation of the wormhole ends. However, this effect falls off exponentially~(\ref{adscoul}).
 In the usual spirit of dilute gas instanton sums, there is not at an exact saddle point of the action but rather a nearly flat plateau parameterized by the positions of the ends.

To make the cluster decomposition argument, consider the gauge theory on a very long periodic Euclidean time $T$, with one set of operators 
${\cal O}_1$ near $\tau = 0$ and another set ${\cal O}_2$ near $\tau = T/2$.  Assuming that the vacuum is unique, we have in the gauge theory that
\begin{equation}
\langle {\cal O}_1 {\cal O}_2 \rangle = \langle 0 | {\cal O}_1 | 0 \rangle  \langle 0 | {\cal O}_2 | 0 \rangle
+ O(e^{-ET/2}) \label{clust}
\end{equation}
where $E$ is the gap to the first excited state.  Possibly in some cases the ground state has a finite degeneracy leading to a finite sum of such terms, but no more than this because of our remarks about the scalar potential.  On the other hand, if the bulk physics is described by $\alpha$-parameters as in Eq.~(\ref{alpha}), we would have the expression
\begin{equation}
\langle {\cal O}_1 {\cal O}_2 \rangle = 
\int d\alpha\, e^{-\frac{1}{2} \alpha_I (C^{-1})_{IJ} \alpha_J}
\langle 0 | {\cal O}_1 | 0 \rangle_\alpha  \langle 0 | {\cal O}_2 | 0 \rangle_\alpha
+ O(e^{-E_\alpha T/2})\ ,
\end{equation}
where the subscript $\alpha$ indicates that quantities are to be calculated using the $\alpha$-shifted action.
This is not equivalent to a product~(\ref{clust}) or a sum of products.\footnote{There is a  large amount of supersymmetry in the bulk, and because this is a gauge symmetry it must be respected by the effective operators induced by the wormhole. These will therefore start at some high dimension, 
but this does not affect the problem of principle with cluster decomposition.}

Thus it appears that quantum gravity as constructed via AdS/CFT duality does not include Euclidean wormholes, or if it does then they do not have the expected effect.
It has been suggested to us by several people that one might get a different gravity, with wormholes, by modifying the CFT.  For example, introducing nonlocal bilinear interactions~(\ref{bilocal}) directly into to CFT would destroy cluster decomposition.  However, this does not seem plausible to us.  Modifications of the CFT correspond to perturbations of the boundary conditions, not the bulk dynamics.  An experimentalist in the bulk should be able to distiguish a local modification of the dynamics from effects propagating inward from the boundary (for example, by doing measurements within a Faraday cage).

The wormhole solutions we have found pose a sharp paradox with AdS/CFT and the apparent 
uniqueness of quantum gravity in maximally supersymmetric backgrounds. One might have hoped that string theory would have avoided such paradoxes by not producing effective field theories allowing wormhole solutions, but that does not appear to be the case.  Instead, these saddle points of the euclidean path integral 
apparently do not contribute to the partition function despite no obvious IR pathologies (beyond the usual ones of euclidean gravity). 
Of course, there is no reason to expect every saddle point of an integral to contribute --- this is already the case even for ordinary integral,s (such as the Airy integral), but one is left to wonder what are the rules that determine saddles contribute and which don't. What is pathological about the wormhole solutions? 

Perhaps there is simply a rule excising topologically non-trivial configuations like wormholes from the approximate sum over geometries.  Or it may be that the fact that the action lies below the BPS action is a clue that these solutions are in a region of field space that is not reached by a proper contour rotation.  The observation of Ref.~\cite{BCPVV}, discussed at the end of the previous section, is a further argument in this direction.

Another interesting observation is that the wormhole solution traverses a large distance in moduli space, in Planck units. 
In analagous situations in Minkowski space, it is difficult to set up backgrounds which 
span super-Planckian ranges in moduli space without generating horizons to shield them. 
Some have taken such arguments to imply that there is no sense in which we can talk about 
distant vacua in moduli space as really part of the same theory, though there have been no convincing arguments on this issue either way. Our wormhole solutions provide a setting where a similar question can be asked. Wormholes exist in the long-distance theory {\it only} when super-Planckian distances are traversed in moduli space.  There are no horizons these excursions can hide behind in euclidean space, and the naive interpretation of wormholes makes connecting distant parts of moduli space in the same theory compulsory. It is therefore interesting that the apparently correct interpretation --- that wormholes don't contribute after all --- also censors this connection. 

Finally, it may be that wormholes do contribute to the path integral but that the interpretation in terms of fluctuating couplings is not correct.  That is, there may be some question in quantum gravity for which these saddles contribute.  See for example Ref.~\cite{Rubakov:1996cn} for an alternate interpretation.  

We thank Juan Maldacena, Rob Myers, Andy Strominger, and Stefan Vandoren for useful discussions.  The work of N. A.-H. is supported by the DOE under contract DE-FG02-91ER40654.  The work of J. O. and J. P. is supported by NSF grants PHY05-51164 and PHY04-56556.


\begin{thebibliography}{99}
   
\bibitem{GS1}
  S.~B.~Giddings and A.~Strominger,
  ``Axion Induced Topology Change In Quantum Gravity And String Theory,''
  Nucl.\ Phys.\ B {\bf 306}, 890 (1988).

\bibitem{LRT}
  G.~V.~Lavrelashvili, V.~A.~Rubakov and P.~G.~Tinyakov,
  ``Disruption of Quantum Coherence Upon a Change in Spatial Topology in Quantum Gravity,''
 JETP Lett.\  {\bf 46}, 167 (1987)
 [Pisma Zh.\ Eksp.\ Teor.\ Fiz.\  {\bf 46}, 134 (1987)].

\bibitem{Hawking}
  S.~W.~Hawking,
  ``Quantum Coherence Down the Wormhole,''
  Phys.\ Lett.\ B {\bf 195}, 337 (1987); \\[3pt]
S.~W.~Hawking,
  ``Wormholes in Spacetime,''
  Phys.\ Rev.\ D {\bf 37}, 904 (1988).
   
\bibitem{Coleman:1988cy}
  S.~R.~Coleman,
  ``Black Holes as Red Herrings: Topological Fluctuations and the Loss of Quantum Coherence,''
  Nucl.\ Phys.\ B {\bf 307}, 867 (1988).
  
\bibitem{GS2}
S.~B.~Giddings and A.~Strominger,
  ``Loss of Incoherence and Determination of Coupling Constants in Quantum Gravity,''
  Nucl.\ Phys.\ B {\bf 307}, 854 (1988).

\bibitem{ccc}
  S.~R.~Coleman,
   ``Why there is nothing rather than something: A theory of the Cosmological Constant''
  Nucl.\ Phys.\ B {\bf 310}, 643 (1988).

\bibitem{Rey:1998yx}
  S.~J.~Rey,
  ``Holographic principle and topology change in string theory,''
  Class.\ Quant.\ Grav.\  {\bf 16}, L37 (1999)
  [arXiv:hep-th/9807241].
  
  \bibitem{Maldacena:2004rf}
  J.~M.~Maldacena and L.~Maoz,
  ``Wormholes in AdS,''
  JHEP {\bf 0402}, 053 (2004)
  [arXiv:hep-th/0401024].

\bibitem{stringwh}
  S.~B.~Giddings and A.~Strominger,
  ``String Wormholes,''
  Phys.\ Lett.\ B {\bf 230}, 46 (1989).

\bibitem{stringwh2}
  K.~Tamvakis,
  ``Two Axion String Wormholes,''
  Phys.\ Lett.\ B {\bf 233}, 107 (1989).

\bibitem{stringwh3}
  E.~Bergshoeff, A.~Collinucci, U.~Gran, D.~Roest and S.~Vandoren,
  ``Non-extremal D-instantons,''
  JHEP {\bf 0410}, 031 (2004)
  [arXiv:hep-th/0406038]; 
``Non-extremal instantons and wormholes in string theory,''
  Fortsch.\ Phys.\  {\bf 53}, 990 (2005)
  [arXiv:hep-th/0412183].
  
\bibitem{sham}
  O.~DeWolfe, A.~Giryavets, S.~Kachru and W.~Taylor,
  ``Type IIA moduli stabilization,''
  JHEP {\bf 0507}, 066 (2005)
  [arXiv:hep-th/0505160].
  
\bibitem{Einhorn:2002am}
  M.~B.~Einhorn,
  ``Instanton of type IIB supergravity in ten dimensions,''
  Phys.\ Rev.\  D {\bf 66}, 105026 (2002)
  [arXiv:hep-th/0201244].
  
\bibitem{Gutperle:2002km}
  M.~Gutperle and W.~Sabra,
  ``Instantons and wormholes in Minkowski and (A)dS spaces,''
  Nucl.\ Phys.\ B {\bf 647}, 344 (2002)
  [arXiv:hep-th/0206153].
  
\bibitem{Coleman:1989ky}
 S.~R.~Coleman and K.~M.~Lee,
  ``Wormholes made without massless matter fields,''
  Nucl.\ Phys.\  B {\bf 329}, 387 (1990).
  
\bibitem{Hawk}
S.~W.~Hawking, in {\it General Relativity, an Einstein Centenary Survey,} eds. S.~W. Hawking and W.
Israel (Cambridge University Press, Cambridge, 1979).

\bibitem{Gibbons:1995vg}
G.~W.~Gibbons, M.~B.~Green and M.~J.~Perry,
  ``Instantons and Seven-Branes in Type IIB Superstring Theory,''
  Phys.\ Lett.\  B {\bf 370}, 37 (1996)
  [arXiv:hep-th/9511080].


\bibitem{BCPVV} 
  E.~Bergshoeff, A.~Collinucci, A.~Ploegh, S.~Vandoren and T.~Van Riet,
  JHEP {\bf 0601}, 061 (2006)
  [arXiv:hep-th/0510048].


  
  
\bibitem{juan}
J.~M.~Maldacena,
  ``The large N limit of superconformal field theories and supergravity,''
  Adv.\ Theor.\ Math.\ Phys.\  {\bf 2}, 231 (1998)
  [Int.\ J.\ Theor.\ Phys.\  {\bf 38}, 1113 (1999)]
  [arXiv:hep-th/9711200].
  
\bibitem{Seiberg:1999xz}
  N.~Seiberg and E.~Witten,
  ``The D1/D5 system and singular CFT,''
  JHEP {\bf 9904}, 017 (1999)
  [arXiv:hep-th/9903224].

\bibitem{GKPholo}
S.~S.~Gubser, I.~R.~Klebanov and A.~M.~Polyakov,
  ``Gauge theory correlators from non-critical string theory,''
  Phys.\ Lett.\ B {\bf 428}, 105 (1998)
  [arXiv:hep-th/9802109].

\bibitem{Wholo}  
E.~Witten,
  ``Anti-de Sitter space and holography,''
  Adv.\ Theor.\ Math.\ Phys.\  {\bf 2}, 253 (1998)
  [arXiv:hep-th/9802150].

\bibitem{Rubakov:1996cn}
  V.~A.~Rubakov and O.~Y.~Shvedov,
  ``A Negative Mode About Euclidean Wormhole,''
  Phys.\ Lett.\  B {\bf 383}, 258 (1996)
  [arXiv:gr-qc/9604038].





\end{thebibliography}
\end{document}